\documentclass{aa}  

\usepackage{graphicx}
\usepackage{txfonts}
\begin{document}

   \title{Star formation in low density H{\sc i} gas around the Elliptical Galaxy NGC~2865}


   \author{F. Urrutia-Viscarra
          \inst{1,2}
          \and
         S. Torres-Flores \inst{1}
         \and
         C. Mendes de Oliveira\inst{3}
         \and 
         E. R. Carrasco\inst{2}
         \and
         D. de Mello\inst{4}
         \and
         M. Arnaboldi\inst{5,6}
          }

   \institute{Departamento de F\'isica y Astronom\'ia, Universidad de La Serena, Av. Cisternas 1200 Norte, La Serena, Chile\\
              \email{furrutia@userena.cl, furrutia@gemini.edu}
              \and
              Gemini Observatory/AURA, Southern Operations Center, Casilla 603 La Serena, Chile
         \and
         Departamento de Astronomia, Instituto de Astronomia, Geof\'isica e Ci\^encias Atmosf\'ericas da USP, Rua do Mat\~ao 1226, Cidade Universit\'aria, 05508-090, S\~ao Paulo, Brazil
         \and
         Observational Cosmology Laboratory, Code 665, Goddard Space Flight Center, Greenbelt, MD 20771, USA
         \and
         European Southern Observatory, Karl-Schwarzschild-Strasse 2, 85748 Garching, Germany
         \and
         INAF, Observatory of Pino Torinese, Turin, Italy }

   \date{Received September 15, 1996; accepted March 16, 1997}

 
  \abstract
   {Interacting galaxies surrounded by H{\sc i} tidal debris are ideal sites
for the study of young clusters and tidal galaxy formation. 
The process that triggers star formation in
the low-density environments outside galaxies is still an open
question. New clusters and galaxies of tidal origin are expected to have high
metallicities for their luminosities.
Spectroscopy of such objects is, however, at the
limit of what can be done with existing 8-10m class telescopes,
which has prevented statistical studies of these objects.}
   {NGC~2865 is an UV-bright merging elliptical galaxy with shells
and extended H{\sc i} tails.
The regions observed in this work were previously detected
using multi-slit imaging spectroscopy.}
   {We obtain new multi-slit spectroscopy of six young star-forming
regions around NGC~2865, to determine their redshifts and metallicities.}
   {The six emission-line regions are located 16-40 kpc from
NGC~2865 and they have similar redshifts.  They have ages of $\sim$
10 Myears and an average metallicity of $\sim$ 12+log(O/H)$\sim$8.6,
suggesting a tidal origin for the regions.  It is noted that they
coincide with an extended H{\sc i} tail,   which has projected density of N$_{HI}$ $<$ 10$^{19}$ cm$^{-2}$, and displays a low surface brightness counterpart.
These regions may represent the youngest of the three populations
of star clusters already identified in NGC~2865.}
 {The high, nearly-solar, oxygen abundances found for
the six regions in the vicinity of NGC~2865 suggest that they were
formed by pre-enriched material from the parent galaxy, from gas
removed during the last major merger.  Given the mass and the
location of the H{\sc ii} regions, we can speculate that these young
star-forming regions are potential precursors of globular clusters
that will be part of the halo of NGC~2865 in the future.  Our result
supports the use of the multi-slit imaging spectroscopy as a useful tool for finding nearly-formed stellar systems around galaxies.}


   \maketitle
%

\section{Introduction}

   Galaxies may experience interactions and mergers throughout their
lifetimes.  Tidal forces distort galaxy shapes leading to the
formation of different structures and substructures, e.g. shells,
rings, tails, and to the onset of star formation inside and outside
galaxies
\citep[e.g.][]{Toomre77,Mendes04,Schiminovich13,Ueda14,Ordenes16}
depending on the nature and evolutionary stage of the on-going tidal interaction. In
particular, after a close encounter of two gas-rich systems of
similar mass, the gas may be stripped from the interacting galaxies
forming long filaments or tidal tails driven by gravity torques
(e.g. the Antennae galaxies, which are the nearest example of merging
disk galaxies in the Toomre 1977 sequence), while the stars mostly remain
in the system, given their higher velocity dispersions and their
collisionless dynamics.  Once the gas has been removed, it can
cool, self-gravitate and form new stars \citep{Duc12}. 
Thus, these systems are ideal laboratories to study star formation
in extreme environments, in particular outside galaxies, in regions
where, under normal conditions, the gas density would have been too
low for star-formation to occur \citep{Maybhate07,Sengupta15}.

The details of the processes capable of triggering star formation
in the low-density environments of galaxies outskirts and in the
intergalactic medium are still not fully understood. Numerical
simulations done to study the dynamics of interacting and merging
galaxies \citep[e.g.][]{Bournaud08,Bournaud10,Escala13,Renaud15} have shown
that young stellar substructures are formed in the outskirts
of merger remnants as well as outside galaxies, in gas clouds 
stipped during interactions.  These simulated
objects have properties similar to the observed ones: the largest objects
are usually formed at the tip of the tails and the objects have low
M/L ratios and high metallicities.  Indeed the actively star-forming
regions associated with the galaxy outskirts or intergalactic medium
\citep[e.g.][]{Neff04,Lisenfeld07,Knierman12,Mullan13} have high
metallicities for their luminosities, given that they are formed
by gas that was pre-enriched in the ``parent" galaxy
\citep[e.g.][]{Mendes04,Duc07,deMello12,Torres12,Torres14}.  The
evolution of the newly formed systems is mainly driven by gravitational
turbulence and instabilities around the Jeans-Scale \citep{Bournaud10}.

Most of the studies on star-forming regions in the outskirts of
galaxies are based in the analysis of ongoing wet mergers, where
H{\sc i}-rich tidal debris and tidal structures are present, and the
interacting galaxies are still separate entities
\citep[e.g.][]{Oosterloo04,Mendes04,Mendes06,Ryan04,Boquien07,Torres12,deMello12,Rodruck16,Lee16}.
However, the environments of peculiar merger-candidate elliptical
galaxies, with H{\sc i} outside their main optical body, have not
receive as much attention.  This is an interesting variation given
that these systems are in advanced stages of evolution.  \cite{Rampazzo07}
 have shown a few examples of ``rejuvenated'' elliptical
galaxies, which display young bursts of star formation.  The object of study in this paper is one of these rejuvenated
ellipticals.  NGC~2865, a genuinely peculiar elliptical galaxy,
with a surface brightness profile consistent with r$^{1/4}$, inside
its effective radius \citep{Jorgensen92}, but deep images show
shells and disturbed morphology, present in merging systems
\citep{Rampazzo07}.  NGC~2865, at a distance of ~ 38 Mpc, has an
extended tidal tail of H{\sc i} gas, settled in a ring around the
galaxy, with low surface brightness optical counterpart.  The fine structures
present around the galaxy are shells, very faint filaments and an
outer loop, that are indicative of an advanced stage of interaction
of $\sim$ 4 Gyr \citep{Malin83,Hau99}.

In \cite{Urrutia14} (hereafter UV14) 
we obtained a complete census of H$\alpha$-emitting sources in the 
southeastern region of the H{\sc i} ring of NGC~2865 using the multi-slit imaging 
spectroscopy (MSIS) technique \citep{Gerhard05,Arnaboldi07}. Using this technique (a combination
of a mask of parallel multiple slits with a narrow-band filter centered
around the H$\alpha$ line, see UV14 for details),
seven candidate intergalactic H{\sc ii} regions were detected. Due to the short wavelength interval
($\sim$ 80 \AA{}) of the spectra, only one emission line (H$\alpha$)
was typically detected. We were, then, not able to confirm the redshifts and compute the metallicities using just one line. Thus, here we
revisited NGC~2865 using multi-object spectroscopy of five of
the seven H{\sc ii} regions previously detected. For one of these
regions we placed two slits over it, and these were resolved into
two different regions. The large wavelength interval used in
the new observations, from 4500 \AA{} to 7000 \AA{} allowed us to confirm
the nature of the sources as newly formed objects and to derive their main
physical parameters.

The paper is organized as follows. In Section 2 we describe the observation and the data reduction.
The analysis and results are described in Section 3. We discuss our results in Section 4 and present our conclusions in Section 5. 

Throughout this paper we use H$_0$ = 75 km s$^{-1}$ Mpc$^{-1}$ which
results in a distance for NGC~2865 of 38 Mpc. At this distance, 1' = 11 kpc. The systemic velocity of NGC~2865 is 2627 km s$^{-1}$

\begin{figure*}
   \centering
   \includegraphics[width=0.7\textwidth]{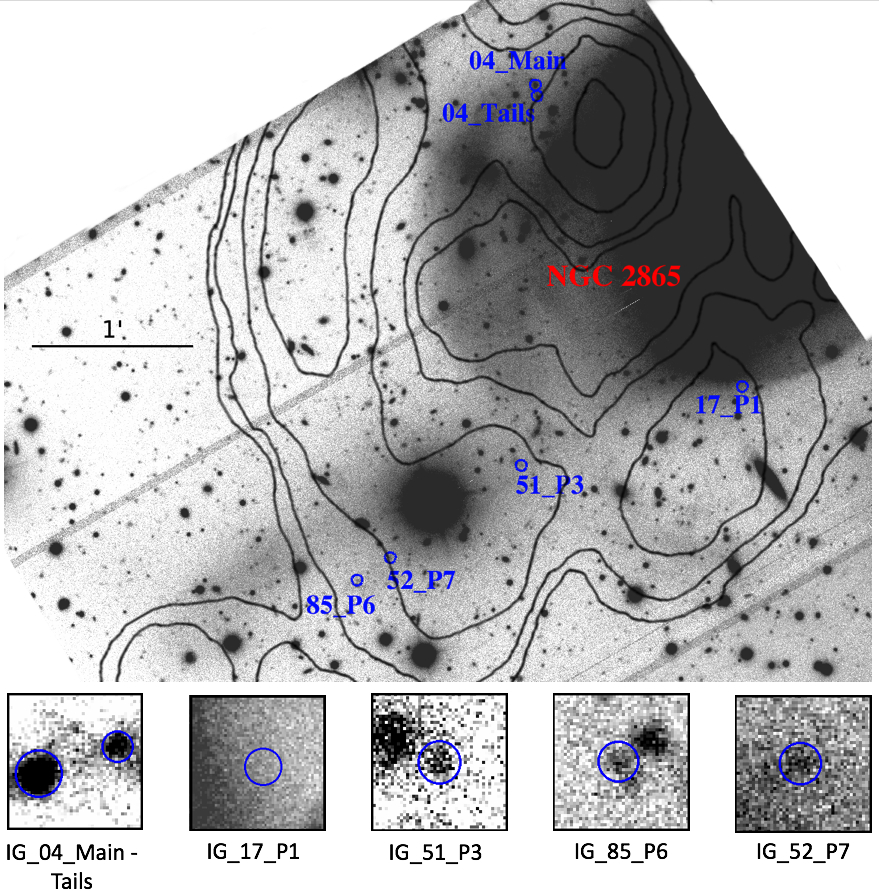}
      \caption{Gemini g'-band images, the regions are marked and a zoom is show for each one. The size of the zoom images are 6.5" $\times$ 6.5". A radius of 1" was used in the photometry, which are indicate in the zoom images. North is up and East is to the left.}
         \label{fig:image}
\end{figure*}

\section{Observation and Reduction}

\subsection{Observation}

The data were collected with the Gemini Multi-Object Spectrograph \citep[][hereafter GMOS]{Hook04} mounted on the Gemini South telescope in Chile in queue mode (Program ID. GS-2011A-Q-55). Given that we wanted to compare the results with those obtained with the MSIS technique, our field of view is the same as observed in UV14. We observed the southeastern H{\sc i} tail of NGC~2865 ($\alpha$(2000) = 9$^h$23$^m$37$\fs$13, $\delta$(2000) = -23${\degr}$11$^m$54$\fs$34) in the g' filter on February 1 2011 (UT) to build the multi-slit mask. 

The spectra were observed between April 13 and April 27 2013, under gray and photometric conditions, and with a seeing ranging between 0\farcs8 and 1\arcsec. We centered the slits in 20 sources across the GMOS field of view, 5 of which were previously identified in UV14. For the region $IG\_04$ we set two slits, one in the stellar cluster (or main source) and the other in the tail, as is defined by UV14 To avoid confusion, we re-defined the ID of the region $IG\_04$ as $IG\_04\_main$ and $IG\_04\_tail$ according to where we set the slit (see Fig. \ref{fig:image}). The spectra in the mask were observed using the R400 grating, 1\arcsec slits, 2 $\times$ 2-binning, and centered at 6550. A total of 12 exposures of 1150 seconds each were obtained. An offset of 50\AA{} towards the blue or the red was performed, between successive exposures, such that the central wavelength ranged from 6400\AA{} to 6650\AA{}, to avoid losing any important emission lines that could fall, by chance, in the gaps between CCDs. Spectroscopic flat fields and CuAr comparison lamps spectra were taken before and after each science exposure. 

 \subsection{Data reduction}

All spectra were reduced with the Gemini GMOS package version 1.8 inside 
IRAF\footnote{{\sc iraf} is distributed by the National Optical Astronomy Observatories,which are operated by the Association of Universities of Research in Astronomy, Inc., under cooperative agreement with the NSF.} following the standard procedures for MOS observations. Science exposures, spectroscopic flats and 
CuAr comparison lamps were overscan/bias-subtracted and trimmed. The two-dimensional science spectra 
were flat fielded, wavelength calibrated, rectified (S-shape distortion corrected) and extracted 
to one-dimensional format. The cosmic rays were removed using the Laplacian Cosmic Ray Identification algorithm \citep{van Dokkum01}. The final spectra have a resolution of $\sim$ 7.0\AA{} (as measured from the sky lines 5577\AA{} and 6300\AA{}) and a dispersion of 1.36 \AA{} pixel$^{-1}$, with a typical coverage in wavelength between $\sim$ 4500\AA{} and $\sim$7000\AA{} (see Fig. 2).

The science spectra were flux calibrated using the spectrum of the spectrophotometric standard star LTT~7379 observed during the night of April 11, 2011 UT, under different observing program and with the same instrument setup used for the science spectra. The standard star being observed in a different night ensures good relative flux calibration, although probably an uncertain absolute zero point. The flux standard was reduced following the same procedures used for the science frames.

\begin{table}
 \centering
 \begin{minipage}{80mm}
  \caption{Physical properties of the intergalactic H{\sc ii} regions.} \label{table:obs}
  \begin{tabular}{@{}lc c c c @{}}
  \hline    
\multicolumn{1}{c}{ID} & $\lambda H\alpha$ & $V_{sys}$ & D$_{projected}$ & Line ratio    \\
 & \AA{} & km s$^{-1}$  & Kpc & ${H\alpha}/{H\beta}$  \\
\hline      
  \hline              
 {\it IG\_04\_main }        &  6618     & 2571	&  16 & 7.27       \\
 {\it IG\_04\_tail }        &  6616     & 2471	&  16 & 6.84       \\
 {\it IG\_17\_P1}           &  6617     & 2461	&  15 &  ---      \\
 {\it IG\_51\_P3}           &  6615	    & 2360 	&  26 & 4.06     \\
 {\it IG\_85\_P6}           &  6614 	& 2341 	&  40 & ---      \\
 {\it IG\_52\_P7}           &  6614 	& 2331 	&  37 & 6.00     \\
\hline
\end{tabular}
\end{minipage}

\end{table}

\subsection{Emission lines}

The six spectra exhibit strong emission lines of H$\alpha$ and [N{\sc ii}]$\lambda$6583\AA{}. Four of the six spectra also show H$\beta$ and the forbidden lines: [O{\sc iii}]$\lambda\lambda$4959, 5007\AA{} and [S{\sc ii}]$\lambda\lambda$6717, 6731\AA{}. No significant underlying continuum was measured in any of the six cases, hence no continuum subtraction was done. The principal parameters for each spectrum presented in Table \ref{table:obs} are: column (1): ID; column (2), the central wavelength for the  H$\alpha$ emission line; column (3), the heliocentric velocity, obtained with the task {\sc rvidline} from {\sc iraf} (using at least three emission lines); column (4), the projected distance (from the center of NGC~2865) in kpc; and column (5), line ratios H$\alpha$/H$\beta$. The latter was used to estimate the the color excess, $E(B-V)$ (see section 3.1). The errors for the velocities listed in column (3) were estimated using Monte Carlo simulations for 100 runs, and were found to be $\sim$ 40 km s$^{-1}$ for each spectrum.

\begin{table*}
 \centering
  \caption{Line intensities of the Intergalactic H{\sc ii} regions around NGC~2865.}  \label{table:lines_flux}
\centering
 \begin{tabular}{l  c c c c c c}     
\hline       
\multicolumn{1}{c}{ID}  & Flux$_{H\beta}$ & Flux$_{[OIII]\lambda5007}$  & Flux$_{H\alpha}$ & Flux$_{[NII]\lambda6583}$ & Flux$_{[SII]\lambda6717}$  & Flux$_{[SII]\lambda6731}$ \\
\hline        \hline            
 {\it IG\_04\_main}            & 0.49 $\pm$ 0.06 & 1.12 $\pm$ 0.36 & 1.27 $\pm$ 0.67  &	0.52 $\pm$ 0.16 & 0.22 $\pm$ 0.03 & 0.20 $\pm$ 0.07  \\
  {\it IG\_04\_tail}            & 2.41 $\pm$ 0.97 & 7.00 $\pm$ 0.76 & 7.63 $\pm$ 1.23  &	2.15 $\pm$ 0.36	& 1.16 $\pm$ 0.35 & 0.69 $\pm$ 0.23  \\
 {\it IG\_17\_P1}    & --- & 0.12 $\pm$ 0.05 & 0.13 $\pm$ 0.02 &  0.03 $\pm$ 0.01 &	---	&  --- \\
 {\it IG\_51\_P3}  & 0.04 $\pm$ 0.01 & --- & 0.18 $\pm$ 0.05  & 0.05 $\pm$ 0.01	& --- & ---  \\
 {\it IG\_85\_P6} & ---  & --- & 0.55 $\pm$ 0.13 & 0.15 $\pm$ 0.03 & 0.09 $\pm$ 0.06 & 0.07 $\pm$ 0.09  \\
 {\it IG\_52\_P7}  & 0.16 $\pm$ 0.35 & 0.10 $\pm$ 0.74 & 0.29 $\pm$ 0.46  & 0.08 $\pm$ 0.02	& 0.06 $\pm$ 0.01 & 0.04 $\pm$ 0.02 \\
\hline                  
\end{tabular} 
\tablefoot{The fluxes are in units of 10$^{-15}$ erg s$^{-1}$ cm$^{-2}$.}
\end{table*}

\section{Analysis and Results}

For each of the six regions, when possible, we derived the following parameters: i) color excess $E(B-V)$, ii) oxygen abundance, 12 + log(O/H), and iii) electron density, $n_e$. These results are used in the following analysis of the physical and chemical properties of the H{\sc ii} regions around NGC~2865.

\subsection{Reddening correction E(B-V)} \label{sec:redd}

Dust extinction in each region was estimated from the line ratio H$\alpha$/H$\beta$ \citep{Calzetti94}. The intrinsic value for the H$\alpha_0$/H$\beta_0$ ratio for an effective temperature of 10$^4$ K and N$_e$ = 10$^2$ cm$^{-3}$ is 2.863 \citep{Osterbrock06}. Thus,  we estimated the extinction using the following relation \citep{Osterbrock06}:

\begin{eqnarray}
\frac{I_\lambda}{I_{H\beta}} & = & \frac{I_{\lambda0}}{I_{H\beta0}}10^{-c[f(\lambda)-f(H\beta)]}
\end{eqnarray}

\noindent where I$_\lambda$ and I$_{\lambda0}$ are the observed and the theoretical fluxes, respectively. c is the reddening coefficient and f($\lambda$) is the reddening curve. The line ratios H$\alpha$/H$\beta$ used to estimate the dust extinction values are tabulated in Table \ref{table:obs}

The relation between $E(B - V)$ and $c$ depends on the shape of the extinction curve; assuming a Galactic standard extinction curve (R = 3.1, typical in the diffuse interstellar medium) and  the H$\alpha$ line ($\lambda$ = 6563\AA{}), the values for each parameter are: f(H$\beta$) = 1.164, f(H$\alpha$) = 0.818. Thus, the color excess is given by $E(B - V) \approx 0.77c$.

For the spectra  {\it IG\_04\_main}, {\it IG\_04\_tail}, {\it IG\_51\_P3} and {\it IG\_52\_P7} the color excess was obtained using the H$\alpha$ and H$\beta$ lines, as described above. Given that H$\beta$ emission is not detected in the spectra of {\it IG\_17\_P1} and  {\it IG\_85\_P6}, we used in these cases a correction equal to the average of the color excess obtained in the other 4 regions. We corrected each spectrum for the interstellar dust extinction using the extinction function given by \cite{Calzetti00}. This is thought to be appropriate for de-reddening the spectrum of a source whose output radiation is dominated by massive stars, as we assume the regions observed here are.

\begin{figure}
   \centering
   \includegraphics[width=\columnwidth]{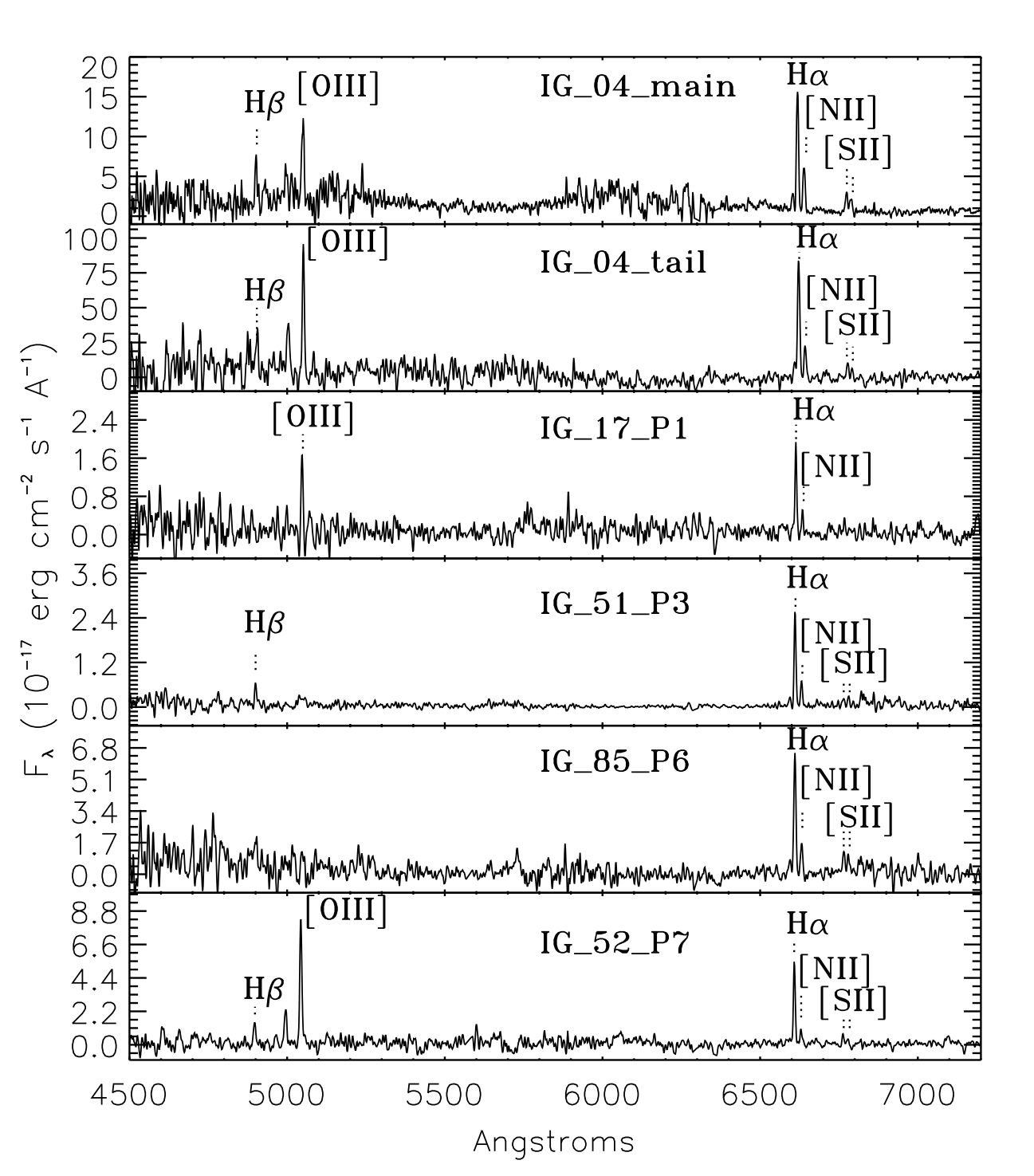}
      \caption{Spectra for 6 intergalactic regions around NGC~2865. Including IG\_04\_tail. }
         \label{fig:spectra}
\end{figure}

The six extracted spectra after the extinction correction are shown in Figure \ref{fig:spectra}. The lines used are marked in each spectrum. In Table \ref{table:lines_flux} we list the resulting fluxes, after the reddening correction is applied.

The four H{\sc ii} regions for which it was possible to estimate the reddening correction, present high values of color excess, E(B-V)$ >$ 0.33. These values indicate a considerably high dust absorption, typical of a nebula where newborn stars are being formed  \citep{Arias06}. Indeed, these four regions display quite young ages, as derived by UV14, ranging from $\sim$ 2 to 50 Myr, which strongly suggests that these sources are the birthplace of young stars. 

\subsection{Oxygen abundance}

The most commonly measured tracers of metallicity are oxygen and iron, given their bright spectral lines. Iron is usually used as a metallicity indicator when old stars are the main component, while oxygen is the metallicity tracer of choice in the interstellar medium \citep{Henry99}.

In the ionized gas the most accurate estimator to derive the chemical abundance is the electron temperature, T$_e$ \citep{Osterbrock06}. In the optical, T$_e$ can be estimated using sensitive lines such as [O{\sc iii}$\lambda$4363\AA{}]. However, the auroral lines are difficult to detect in extragalactic star forming regions, where cooling is efficient and the temperature is low. Thus, strong nebular lines must be used to derive oxygen abundances \citep[e.g.][]{Denicolo02,Pilyugin01,Pettini04}. The empirical methods more frequently use the N$_2$ and O$_3$N$_2$ indices,

\begin{equation}
N_2 \equiv log \left( \frac{I([NII]\lambda 6583)}{I(H\alpha)} \right)
\end{equation}

 \begin{equation}
O_3N_2 \equiv log\left(  \frac{I([OIII]\lambda 5007)}{I(H\beta)} \Big/  \frac{I([NII]\lambda 6583)}{I(H\alpha)} \right) 
\end{equation}

\noindent We calculated the oxygen abundances using the empirical methods N$_2$ and, when possible, also O$_3$N$_2$, as proposed and calibrated by \cite{Pettini04}. These ``empirical'' methods are adequate for estimating oxygen abundances in extragalactic H{\sc ii} regions. We used the linear relation between the oxygen abundance and the N$_2$ index given by:

\begin{equation}
12 + log(O/H) = 8.90 + 0.57 \times N_2 
\end{equation}

\noindent Three regions have the four lines; for these we were able to use the index O$_3$N$_2$ and the linear relation between the O$_3$N$_2$ and the oxygen abundance given by:

\begin{equation}
12 + log(O/H) = 8.73 - 0.32 \times O_3N_2
\end{equation}

\noindent The uncertainties in the calibration of these methods are 0.18 dex and 0.14 dex (at the 68$\%$ confidence interval), for N$_2$ and O$_3$N$_2$, respectively. Table \ref{table:result} lists the oxygen abundances estimated including the values of the N$_2$ and O$_3$N$_2$ indices for each region.  
 
 Table \ref{table:result} summarizes the oxygen abundances calculated for all of our H{\sc ii} regions. using the N$_2$ index. We found an excellent agreement between the values derived from the N$_2$ and the O$_3$N$_2$ indices. The average value for the oxygen abundance for the six regions is 12+log(O/H) $\sim$ 8.6, which is very similar to typical values for extragalactic H{\sc ii} regions reported in the literature for other interacting systems \citep[e.g.][]{Werk10,Torres12,deMello12,Olave15}.

\subsection{Electron density}

The electron density, N$_e$,  is one of the fundamental physical parameters used to characterize H{\sc ii} regions. It is a reliable physical quantity that is not sensitive to flux calibration errors and reddening. It is largely independent of metallicity, and only weakly depends on electron temperature. The most used emission lines to measure N$_e$ are [O{\sc ii}]$\lambda$3729\AA{}/$\lambda$3726\AA{}, and [S{\sc ii}]$\lambda$6716\AA{}/$\lambda$6731\AA{}. Given the wavelength range of our spectra,  N$_e$ was determined for four H{\sc ii} regions based on the measured intensity ratio [S{\sc ii}]$\lambda6716\AA{}$/ $\lambda6731\AA{}$. We used the task {\sc temden}, available in the {\sc nebular} package of {\sc stsdas}, inside {\sc iraf V2.16}. For the calculation we assume an electron temperature of 10$^4$ K, which is a representative value for H{\sc ii} regions \citep{Osterbrock06}. This assumption was necessary, given that we have no independent measurements for the   temperatures of these regions.

It was possible to obtain the electron density for four of the six H{\sc ii} regions, see Table \ref{table:result}. We estimated values which ranged from 90 to 327 cm$^{-3}$, which are typical values of the electron density observed in the Magallanic Clouds \citep{Feast64}, and in particular for bright regions in 30 Dorado \citep{Osterbrock06}. \\

Table \ref{table:result} lists the physical parameters described above using the [O{\sc iii}], H$\beta$, [N{\sc ii}], H$\alpha$ and [S{\sc ii}] lines of the intergalactic H{\sc ii} regions. All parameters were obtained after applying reddening correction. 

\begin{table*}
 \centering
  \caption{Physical parameters derived from spectral lines for the intergalactic H{\sc ii} regions.}\label{table:result}
  \begin{tabular}{@{}l c c c c c c  c@{}}
  \hline    
\multicolumn{1}{c}{ID} & M$_{g'}$&N$_e$ &  E(B-V) & N$_2$ & O$_3$N$_2$ &  \multicolumn{2}{c}{12+log(O/H)}   \\
 & mag & cm$^{-3}$ &  &     &  & N$_2$  & O$_3$N$_2$   \\
\hline     \hline                
 {\it IG\_04\_main}       &	-15.76	& 327   & 0.91	&  -0.39				& 0.74	& 8.67 $\pm$ 0.18 & 8.49 $\pm$ 0.1\\
 {\it IG\_04\_tail}     &	-15.31	&  90  & 0.84	&-0.54 				&1.01 	& 8.58 $\pm$ 0.18 & 8.40 $\pm$ 0.14\\
 {\it IG\_17\_P1} &	---	&    ---  	& 0.70$^*$	&  -0.58				&--- 	& 8.56 $\pm$ 0.18 & --- \\
 {\it IG\_51\_P3} &	-14.58	&  --- 	& 0.33 	&  -0.54   			& --- 	& 8.58 $\pm$ 0.18 & ---  \\
 {\it IG\_85\_P6} & -14.22		& 176 	& 0.70$^*$  	&  -0.54 			& --- 	& 8.58 $\pm$ 0.18 & ---  \\
 {\it IG\_52\_P7} &	-14.51	&  176  	& 0.75 	&  -0.65 			& 1.29 	& 8.52 $\pm$ 0.18 & 8.31 $\pm$ 0.14 \\
\hline               
\end{tabular}
\tablefoot{* Given that the H$\beta$ emission line was not detected in two cases, the color excess, E(B-V), was assumed to be an average of the values derived for the other four spectra}
     \end{table*}   
 
\subsection{Confirming the MSIS technique as an effective tool to find emission-line regions}

The MSIS technique has been successfully used to detect extragalactic
planetary nebulae in the Coma and Hydra {\sc i} clusters
\citep[e.g.][]{Gerhard05,Ventimiglia11}. This technique combines a
mask of parallel multiple slits with a narrow-band filter, centered
around the [O{\sc iii}] $\lambda$5007\AA{} line, at the redshift of the
object.  In the case of the Coma cluster, \cite{Gerhard05} were able to
detect faint objects with fluxes as low as 10$^{-18}$ erg s$^{-1}$
cm$^{-2}$ \AA$^{-1}$. An attempt to use the same technique to detect
star-forming regions in the outskirts of galaxies was done by UV14, for the case of NGC~2865. Given the higher contrast
between emission and background, this technique has great potential
to detect faint emission-line objects when regular narrow-band
imaging would fail to lead to a detection of faint objects.  However,
given that the spectral range observed with such a technique is
limited, the spectra shown in this paper were a necessary follow-up
to validate the efficiency of the MSIS method.

In the present paper we find that all of the detected objects with
MSIS technique (UV14) for which further spectroscopic
follow-up was done, were confirmed to be star-forming regions.  This
demonstrates that the MSIS technique is very efficient in finding
emission-line objects.

This type of technique will become increasingly useful as the new
generation of extremely large telescopes comes on-line and blind
searches for emission-line objects can be efficiently done.

\section{Discussion}

In the previous sections, we described the main properties of the
six intergalactic H{\sc ii} regions found in the southeast ring of
H{\sc i} gas within a radius of $\sim$ 50 kpc of the elliptical
galaxy NGC~2865. The study is based on 6 emission lines: H$\beta$,
[O{\sc iii}]$\lambda$ 5007\AA{}, H$\alpha$, [N{\sc ii}], and S{\sc
ii}$\lambda\lambda$ 6717,6731\AA{}. The six regions were previously detected using
the MSIS technique by UV14, but only one line (H$\alpha$)
was observed then.  The detection of at least three emission
lines in each of the six regions, shown in this work, indicate that
the sources  have a velocity difference of $\sim$ 50 - 300 km s$^{-1}$ with respect to the systemic velocity of NGC~2865 and
they have close to solar oxygen abundance  \citep[12 + log (O/H) $\approx$ 8.6,][]{Asplund09}.

\subsection{Star formation outside galaxies}

The six intergalactic H{\sc ii} regions are located 
16-40 kpc from the center of NGC~2865 and they show strong emission in H$\alpha$ and low or absent continuum.
The indication that they are young come from the ages derived in UV14, from ultraviolet fluxes, but our new observations
constrain the ages further: high emission in H$\alpha$ is a direct
indicator of recent star formation, showing that the last starburst
occurred less than 10 Myr ago, assuming single stellar populations \citep{Leitherer99}.


Indication that they are formed in situ come from the prohibitively
large velocities derived for these regions under the assumption
that they traveled from inside the parent galaxy to the present
location.  Approximate velocities can be derived in back-of-envelope
calculations, taking as an input the projected distance of the
objects (with respect to the parent galaxy) and the derived ages.
If the intergalactic H{\sc ii} regions were thrown out of the disk of the
main galaxy during an interaction, the typical velocities they would
need to have is over 16 kpc / 10 Myr $\approx$ 1600 km s$^{-1}$, which is not expected, given the low velocity dispersions present in the field or in groups of galaxies (typically 200 km s$^{-1}$, arguing in favor of in-situ formation.

In
addition, their metallicities were found to be close to solar.
These high metallicities could be explained if the regions were
formed out of enriched gas which was removed from the galaxy during
the merger event. Thus, these regions are similar to extragalactic
H{\sc ii} regions found in recent or ongoing mergers
\citep[e.g.][]{Mendes01,deMello12,Olave15}.  \cite{Torres12} found
similar objects around the system NGC~2782, an ongoing merging
galaxy which shows optical and H{\sc i} tidal tails containing several
H{\sc ii} regions. These objects have ages of ~1 to 11 Myr,  masses
ranging from 0.8 - 4 $\times$ 10$^{4}$ M$_\odot$ and solar metallicities,
similar properties to those of the regions found in the present
work.  

Other examples of star formation outside galaxies is provided by the intra-cluster compact H{\sc ii} regions detected in the Virgo core, 
close to NGC 4388 \citep{Gerhard02}. These  objects have ages of about ~3 Myr, mass of the order of 4$\times$10$^2$ M$_\odot$ and metallicity 0.25 solar.

What makes NGC~2865 a special target for searches of newly-formed star forming
objects is its evolutionary stage.
This galaxy is a merger relic,
an elliptical galaxy with an r$^{1/4}$ profile and low surface brightness
tidal tails. Its last major merger happened $\sim$
4 Gyr ago \citep{Schiminovich95}. Nevertheless, NGC~2865, in such
an advanced stage of evolution, proved to contain
star forming regions with very similar properties to those found 
in less evolved systems.

\subsection{The faint-continuum sources in low N$_{HI}$ regions}

The regions measured in this work have weak emission in the far- and near- ultraviolet  ($\sim$ log(FUV) $\sim$ 37, $\sim$ log(NUV) $\sim$ 36 erg s$^{-1}$ \AA$^{-1}$) and mass of Log(M$_\star$) $\sim$ 6 M$_\odot$. Five of the six regions have low fluxes in the g'-band, M$_{g'}$ $>$ 15.76 and one of them is not seen at all in the Gemini/GMOS images (i.e. which sets a lower limit of 21 mag in g' for the magnitude of the object).
Regions with such faint optical counterparts can only be found in blind searches like those done using the MSIS technique.

 \cite{Maybhate07} and \cite{Mullan13} reported that the threshold $N_{HI}$ value to form stars in tidal debris of interacting galaxies is log($N_{HI}$) $\approx$ 20.6 cm$^{-2}$, which is considered a necessary but not a sufficient condition to generate clusters. In the case of NGC~2865, this value is one order of magnitude lower with H{\sc i} density of $\sim$ 3.8 $\times$ 10$^{19}$ cm$^{-2}$, although it is possible that the real H{\sc i} column densities are underestimated due to the large size of the beam \citep{Schiminovich95}. The level of star formation in the southeast H{\sc i} cloud where the regions of NGC~2865 are located is 2.6 $\times$ 10$^{-3}$ M$_\odot$ yr$^{-1}$ (UV14), which is similar to values observed in other systems such as in the Leo ring and in NGC~4262 \citep{Thilker09,Bettoni10}. 
 
 The mechanisms that trigger and quench star formation in such low-gas density environments is still a mistery. However, what can be said is that star formation does happen outside galaxies, in low H{\sc i} column density regions, in different systems, including around at least one merger relic, NGC~2865.

\subsection{Are H{\sc i} tidal debris and tails nurseries of future globular clusters?}

NGC~2865 is a shell galaxy \citep{Malin83} with a peculiar morphology,
embedded in a ring-shape H{\sc i} tidal tail \citep[see Fig. \ref{fig:image} and also][]{Schiminovich95}. These features all point to a formation
scenario for the galaxy as a wet merger.  Using stellar spectroscopy
and {\it UBV} images, \cite{Schiminovich95} proposed that a possible
major merger event happened between 1 and 4 Gyr ago.  \cite{Rampazzo07} found a young stellar sub-population in this galaxy with an age of 1.8 $\pm$ 0.5 Gyr. This finding is in agreement with the results obtained by \cite{Hau99}, who derived burst ages from population synthesis of 0.4 - 1.7 Gyr, indicating the presence of a younger stellar population in the core of NGC~2865. These authors
also derived a much younger age for the shells of this galaxy.  \cite{Salinas15} and
\cite{Sikkema06} found a very blue population of globular clusters (GCs)
with a color distribution which peaked at ({\it V - I}) = 0.7. This
suggests the presence of stellar populations with ages ranging
from 0.5 to 1 Gyr, similar to those found by \cite{Hau99} in the central
regions of NGC~2865.
Moreover, \cite{Trancho14} found a sub-population of
young metal-rich GCs of age $\sim$ 1.8 $\pm$ 0.8 Gyr. 

In this paper we found
a much younger sub-population, formed by emission-line regions,
with an age $<$ 10 Myr (as indicated by the H$\alpha$-line emission).
All these young regions are located in the H{\sc i} tidal tails and
they could be the result of a more recent burst of star formation. In UV14, we estimated a mass for these regions between 4 $\times$ 10$^4$ and 5 $\times$ 10$^6$
M$_\odot$. Similar values were found by \cite{Maraston04} for the
mass of a young cluster in the merger remnant NGC~7252 and by
\cite{Goudfrooij01} for the masses of clusters around the peculiar
central galaxy of the Fornax cluster, NGC~1316, which is also
considered a merger remnant.

In conclusion, NGC~2865 contains at least three generations of star
clusters, the youngest generation (which is embebed in its H{\sc ii} regions) is the object of study of this
paper.  We do not exclude the possibility that these young H{\sc
ii} regions could be the progenitors of the  intermediate age or the red globular clusters that will be
part of the halo of the future relaxed elliptical galaxy.

\section{Conclusions}

We have confirmed the tidal origin of six H{\sc ii} regions in the
immediate vicinity of NGC~2865, previously detected using the MSIS
technique.  Our results show that although NGC~2865 has undergone
a major merger event $\sim$ 4 Gyr ago, stars are still being formed
in its H{\sc i} tidal tail, outside the main body of the galaxy.

The H{\sc ii} regions around NGC~2865 display high oxygen abundances
of 12+log(O/H)$\sim$8.6. This suggests that they were formed from
pre-processed and enriched material from the parent galaxy which
was stripped to the intergalactic medium during a merger event. The
objects coincide with a gaseous tail with a projected H{\sc i}
density of N$_{HI}$ $<$ 10$^{19}$ cm$^{-2}$, which is one magnitude
lower than the H{\sc i} critical density for star formation, as
given by previous studies. The young regions ($< 10 Myr$) found
in this work can be associated with  
 the youngest population in addition to the two populations of star clusters already identified in NGC 2865:
  i) A much older generation of globular  clusters, and ii) a secondary population with an age ranging from 0.5  to 1.8 Gyrs, as found by \cite{Sikkema06} and \cite{Trancho14}.

 The fate of these young sources is unclear: they may either get re-acreted
onto the parent galaxy, get dissolved or merge to form  massive
(globular) star clusters.  Indeed, given the mass and the location
of the H{\sc ii} regions, we can not exclude that these young star-forming
regions are potential precursors of globular clusters that will be
part of the halo of NGC~2865. The latter scenario is consistent with the results obtained by \cite{Bournaud08}, who used high-resolution simulations of galaxies mergers. These authors found that super star cluster formed in mergers are likely the progenitors of globular clusters. In this sense, our observations are in agreement with the predictions derived from simulations.

Finally, in this paper we verify that the MSIS technique is a
powerful tool to detect faint objects with emission lines which are
not always detected in broad band imaging. This technique may become
specially useful in blind searches of emission line objects to be
done with the next generation of extremely large telescopes.

\section*{Acknowledgments}

The authors would like to thank the anonymous referee for the thoughtful comments which improved the clarity of this paper. Based on observations obtained at the Gemini Observatory, which is operated by the Association of Universities for Research in Astronomy, Inc., under a cooperative agreement with the NSF on behalf of the Gemini partnership: the National Science Foundation (United States), the National Research Council (Canada), CONICYT (Chile), Ministerio de Ciencia, Tecnolog\'{i}a e Innovaci\'{o}n Productiva (Argentina), and Minist\'{e}rio da Ci\^{e}ncia, Tecnologia e Inova\c{c}\~{a}o (Brazil). F.U.-V. acknowledges the financial support of the Chilean agency Conicyt $+$ PAI/Concurso nacional apoyo al retorno de investigadores/as desde el extranjero, convocatoria 2014, under de contract 82140065. ST-F acknowledges the financial support of Direcci\'on de Investigaci\'on y Desarrollo de la ULS, through a project DIULS Regular, under contract PR16143. CMdO acknowledges support from FAPESP  (grants 2009/54202-8, 2016/17119-9  ) and CNPq (grant 312333/2014-5). STF and CMdO kindly thank CONICYT for funding provided through the PAI MEC program for a visit of CMdO to the U. of La Serena.

\end{document}